\documentclass[twocolumn,prl,aps,nobalancelastpage,amsfonts,citeautoscript]{revtex4}

\usepackage{graphicx}
\usepackage{bm}

\begin{document}

\title{Specific heat measurements of the gap structure of the organic superconductors $\kappa-$(ET)$_2$Cu[N(CN)$_2$]Br and $\kappa-$(ET)$_2$Cu(NCS)$_2$}
\author{O.J. Taylor,$^1$ A. Carrington,$^1$ and J.A. Schlueter$^2$ }
\affiliation{$^1$ H.H. Wills Physics Laboratory, University of Bristol, Tyndall Avenue, Bristol,
United Kingdom. \\
$^2$ Materials Science Division, Argonne National Laboratory, Argonne, Illinois 60439}

\date{\today}

\begin{abstract}
We present high resolution heat capacity measurements for the organic superconductors
$\kappa-$(ET)$_2$Cu[N(CN)$_2$]Br and $\kappa-$(ET)$_2$Cu(NCS)$_2$ in fields up to 14~T. We use the
high field data to determine the normal state specific heat and hence extract the behavior of the
electronic specific heat $C_{el}$ in the superconducting state in zero and finite fields. We find
that in both materials for $T/T_c\lesssim 0.3$, $C_{el}(H=0)\sim T^2$ indicating $d$-wave
superconductivity. Our data are inconsistent with $s$-wave behavior, but may be fitted to a strong
coupling $d$-wave model over the full temperature range.
\end{abstract}
\maketitle

The organic superconductors, $\kappa-$(ET)$_2$X, have many similarities to the high temperature
cuprate superconductors (HTSC) \cite{Mckenzie97}. In both cases, the electronic structure is
quasi-two-dimensional and the superconducting phase emerges from an antiferromagnetic insulating
(AFI) state as the phase diagram is transversed. In the case of the cuprates, the structure is
`tuned' away from the AFI state either by varying the oxygen content or making non-isovalent
substitutions, whereas in $\kappa-$(ET)$_2$X it is achieved either by changing the anion X or by
applying external pressure. It is natural then to speculate that mechanism for superconductivity in
these two materials may be related even though $T_c$ is up to one order of magnitude higher in
HTSC.

A first step towards determining if this is indeed the case is to determine the symmetry of the
superconducting energy gap functions in the two families of materials.  The case of the cuprates
has been very well studied and the overwhelming consensus is that these materials have a gap with
predominately $d_{{x^2}-{y^2}}$ symmetry \cite{Vanharlingen95}.  In the organic materials the
situation is more controversial \cite{Lang03}.

The two most widely studied materials are $\kappa-$(ET)$_2$Cu[N(CN)$_2$]Br and
$\kappa-$(ET)$_2$Cu(NCS)$_2$ (hereafter abbreviated to $\kappa$-Br and $\kappa$-NCS) as these have
the highest $T_c$ at ambient pressure ($\sim$12K and $\sim$9.5K respectively). As yet, no direct
phase sensitive determinations of the gap function have been reported, however, there have been
numerous experiments which have probed its anisotropy. Some early experimental determinations of
the temperature dependence of the penetration depth $\lambda(T)$ reported behavior which was
consistent with a $d$-wave gap whereas others were consistent with $s$-wave \cite{Lang03}.  The
discrepancies stem from experiments not having been carried out at sufficiently low temperature
and/or with high enough precision to be conclusive. More recent measurements performed down to
$T/T_c \lesssim 0.03$ showed clearly the existence of low energy excitations which were consistent
with a $d$-wave gap \cite{LeLSWUBRSSYWKCW92,CarringtonBPGKSWGW99}. Thermal conductivity $\kappa(T)$
data shows a low temperature $T$-linear term \cite{BelinBD98} and a four fold variation with basal
plane angle in an applied magnetic field \cite{IzawaYSM02}. Both of these results indicate a
$d$-wave gap as do tunnelling \cite{AraiINTYNA01} and NMR experiments \cite{MayaffreWJLB95}.

Specific heat $C$ has an advantage over many other techniques in that it is a bulk thermodynamic
probe. For example, it is largely insensitive to surface contamination (or thin layers of damaged
material) which may adversely affect probes such as $\lambda(T)$ (in the Meissner state) or
tunnelling. A disadvantage is that the electronic component $C_{el}$ is often only a few percent of
the total at $T_c$. It is difficult to accurately extract $C_{el}$ from the total which is
dominated by phonon contributions.  In Ref.~\cite{NakazawaK97}  $C_{el}$ of $\kappa$-Br was
determined by subtracting an estimate of the phonon contribution obtained from a
non-superconducting quench cooled deuterated version of the same compound.   These authors found
that $C_{el}\sim T^2$ as expected for a $d$-wave gap. However, this approach was criticized by
Elsinger \textit{et al.} \cite{ElsingerWWHSS00} and M\"{u}ller \textit{et al.} \cite{MullerLHSS02}
who instead determined the phonon contribution by applying a high magnetic field to destroy the
superconductivity. These authors claimed that their data for both $\kappa$-Br and $\kappa$-NCS were
well described by an $s$-wave gap. It should be mentioned however, that in neither of these two
reports were attempts made to explicitly fit their data to a $d$-wave model.

Here we report high resolution measurements of $C_{el}$ for $\kappa$-Br and $\kappa$-NCS which are
well described by a strong coupling $d$-wave form of the superconducting gap.  The low temperature
behavior is inconsistent with an $s$-wave gap.

Samples of both compounds were grown by the usual electrochemical method \cite{KiniGWCWKVTSJW90} in
Argonne,  and had masses in the range of 80-600 $\mu$g.  Specific heat measurements were conducted
in a purpose built calorimeter which uses a long relaxation method similar to that described in
Ref.~\cite{WangPJ01}.  Briefly, a Cernox \cite{Lakeshore} chip resistor (CX-1030-Br) is suspended
by silver coated glass fibers in vacuum. The Cernox material acts as both thermometer and heater.
The sample was attached to the calorimeter chip with Apiezon N grease. The addenda (chip, grease,
and leads) was determined in a separate run immediately prior to the main experiment.  The
thermometer was calibrated in field, in 1T increments up to 14T, by stabilizing the temperature
with a capacitance thermometer. The performance of the experiment was extensively studied by
measurement of high purity samples of Ag (with masses in the range 0.3-5mg). In the range 1.3K to
20~K the absolute values of the Ag data agreed with standard values to within 1\%. Small ($<$4mK)
adjustments to the calibration points were made to ensure that the measured $C$ for all Ag samples
were smooth and field independent within experimental error.

\begin{figure}
\center
\includegraphics*[width=8cm]{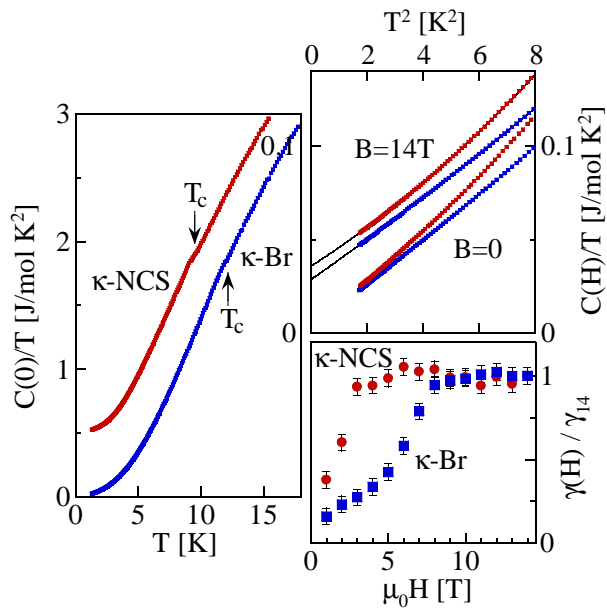}
\caption{(Color online) Left panel: Zero field specific heat data for $\kappa$-Br (abbreviated Br)
and $\kappa$-NCS (abbreviated NCS). Upper right Panel: $C/T$ versus $T^2$ for for both compounds in
fields of 0 and 14T.  The upper curve for each field is $\kappa$-NCS. Lower right panel: Field
dependence of $\gamma$ for both compounds.} \label{figraw}
\end{figure}

Heat capacity data for both materials is shown in Fig.~\ref{figraw}.  In this raw data the
$\sim$3\% anomaly at $T_c$ is barely discernable.   In order to subtract the large phonon
contribution we have made measurements in magnetic fields up to 14T, applied perpendicular to the
basal plane. For both materials, the maximum field is significantly in excess of the upper critical
field $H_{c2}$, and so at 14T both materials are in the normal state.  In Fig.~\ref{figraw} we show
the low temperature portion of the 14T data plotted as $C/T$ versus $T^2$. Fitting this with a
second order polynomial, we determine the Sommerfeld coefficient $\gamma$ as well as the
coefficients of the leading  phonon terms, $\beta_3$ and $\beta_5$ ($C=\gamma T+ \beta_3 T^3 +
\beta_5 T^5$).  The field dependence of $\gamma$ is shown in Fig.~\ref{figraw} and is seen to
saturate at $\gamma=28\pm2$ mJ/mol K$^2$ for $\mu_0H\gtrsim 8$T in $\kappa$-Br and $\gamma=35\pm2$
mJ/mol K$^2$ for $\mu_0H\gtrsim 3$T  in $\kappa$-NCS.   In what follows we make the assumption that
the 14T data is equivalent to that of the normal state in zero field (i.e., the only field
dependence in $C$ is due to the superconductivity).  In principle, there could be magnetic
contributions which vary with field. However, the insensitivity of $C$ to $H$ at high fields
indicates that these contributions are negligible.  Indeed, in Ref.~\cite{NakazawaTKK00} a sizable
magnetic contribution in high field was only found for $T$ well below 1K.

\begin{figure}[t]
\center
\includegraphics*[width=7cm]{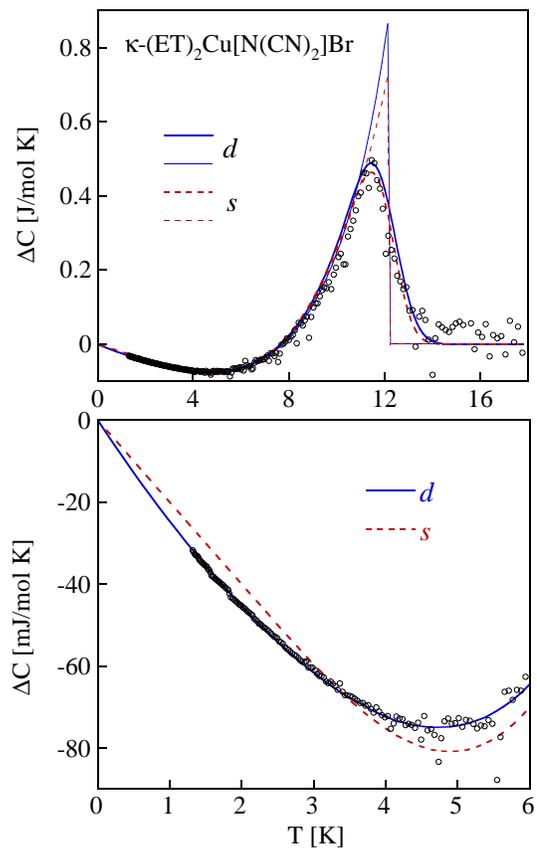}
\caption{(Color online) Top panel: $\Delta C = C(0)-C(14\rm{T})$ versus $T$ for $\kappa$-Br, along
with several fits. The thin solid line is a fit to the strong coupling $d$-wave model, whereas the
thick solid line is the same fit convoluted with a Gaussian. Similarly the dashed lines are fits to
the strong coupling $s$-wave model.  Bottom panel: Enlarged  view  of the low temperature part of
the upper panel (the convoluted fits are indistinguishable and are omitted for clarity).}
\label{figCBr}
\end{figure}

\begin{figure}[t]
\center
\includegraphics*[width=7cm]{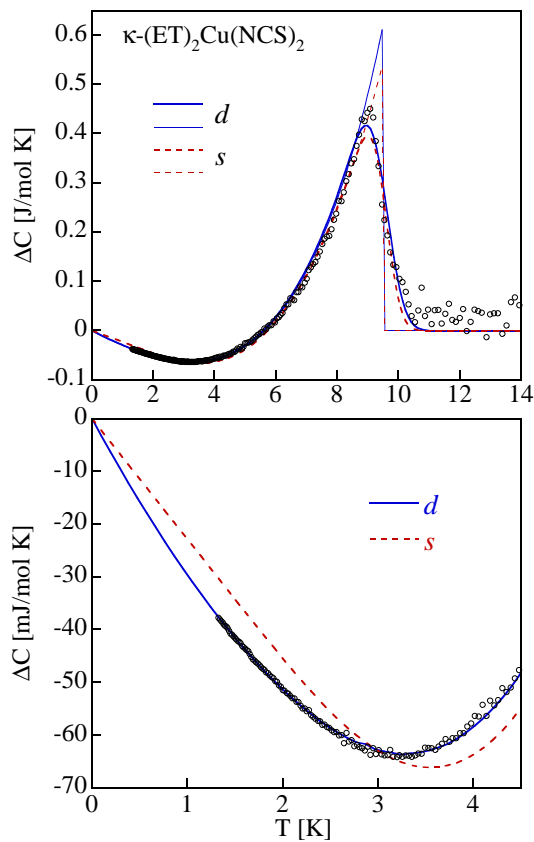}
\caption{(Color online) The same plots as Fig.~\ref{figCBr} but for $\kappa$-NCS} \label{figCNCS}
\end{figure}

In Figs.~\ref{figCBr} and \ref{figCNCS} we show $\Delta C = C(0)-C(14\rm{T})=C_{el}-\gamma T$ for
$\kappa$-Br and $\kappa$-NCS respectively.  The superconducting transitions are now clearly
visible, with the midpoint of the transition giving $T_c$=12.25~K and 9.56~K respectively. Note
that the noise in the data is larger at higher temperature because the fractional resolution of the
calorimeter $\Delta C/C$ is roughly constant with $T$ whereas the phonon background increases $\sim
T^3$.

In many superconductors, the weak coupling form of the BCS theory is inadequate to describe in
detail the physical properties. A full solution to the strong coupling theory is complicated, and
dependent on microscopic details, but it is found that many properties can be explained
satisfactorily with the so-called $\alpha$-model \cite{PadamseeNS73}. Here, the temperature
dependence of the energy gap $\Delta$ is approximated by the weak-coupling behavior but the value
at zero temperature is an adjustable parameter. This model has been used to describe a wide range
of superconductors including  `exotic' materials such as MgB$_2$ \cite{BouquetWFHJJP01} and
NbSe$_2$ \cite{FletcherCDRBPOG07}.

Within the $\alpha$-model, the entropy $S$ in the superconducting state for a two dimensional
cylindrical Fermi surface, is given by
\begin{equation}
\frac{S}{\gamma_n T_c} = \frac{3}{\pi^3}\int_0^{2\pi}\int_0^\infty f\ln f + [1-f]\ln[1-f]
d\varepsilon d\phi
\end{equation}
where the Fermi function $f=[\exp(E/k_BT)+1]^{-1}$, the quasiparticle energy
$E^2=\varepsilon^2+\Delta^2(\phi)$, $\gamma_n$ is the normal state $\gamma$ and the energy gap
$\Delta$ is function of the in-plane angle $\phi$.  The specific heat $C_{el}=T\frac{\partial
S}{\partial T}$.  For conventional isotropic $s$-wave superconductivity the gap function
$\Delta(\phi,T)=\alpha \Delta^s_{\rm BCS}(T)$, whereas in the simplest case for $d$-wave
$\Delta(\phi)=\alpha \Delta^d_{\rm BCS}(T)\cos 2 \phi$. In these expressions $\Delta^{s,d}_{\rm
BCS}$ takes the usual $s$ or $d$-wave weak coupling form.

\begin{table}
\begin{center}
\caption{Parameters derived from the $s$ and $d$ wave fits to the data in Figs.~\ref{figCBr} and
\ref{figCNCS}.  The units of $\gamma$ are mJ/mol K$^2$. $\gamma_{14}$ is the value of $\gamma_n$
derived from a fit to the 14T data. The maximum gaps at zero temperature  $\Delta_0=2.14 \alpha
k_BT_c$ for the $d$-wave fits.} \label{fitparams}
\begin{tabular}{rcccccc}
\hline\hline
&&&\multicolumn{2}{c}{$d$-wave}&\multicolumn{2}{c}{$s$-wave}\\
&$T_c$&$\gamma_{14}$&$\alpha$&$\gamma_n$&$\alpha$&$\gamma_n$\\
$\kappa$-Br&12.25K&28$\pm$2&1.73&26.6&1.47&20.0\\
$\kappa$-NCS&9.56K&35$\pm$2&1.45&33.3&1.34&22.8\\
\hline \hline
\end{tabular}
\end{center}
\end{table}

To allow for the possibility of any part of the sample being non-superconducting (and metallic) we
allow $\gamma_n$ to vary in the fit, so the free parameters are $\alpha$, $\gamma_n$ and $T_c$. As
can be seen in the top panels of Figs.\ \ref{figCBr} and \ref{figCNCS}, at high temperatures the
fits to the $s$ and $d$ models are virtually indistinguishable and both fit the data very well.
Close to $T_c$ the superconducting transition is broadened by inhomogeneity and fluctuation effects
and the fit is considerably improved by convolution with a Gaussian (of width $\sigma$ = 0.65~K and
0.43~K for $\kappa$-Br and $\kappa$-NCS respectively).

At lower temperature (bottom panels of Figs.~\ref{figCBr} and \ref{figCNCS}) there is a very
significant difference between the two models. The $d$-wave model fits the data almost exactly over
the full temperature range whereas the $s$-wave model completely fails at low temperature.  The
parameters derived from the fits are given in Table \ref{fitparams}.  In both materials, for the
$d$-wave fit $\gamma_n$ is found to be very close to the value found from the direct fit to the 14T
data $\gamma_{14}$, whereas the $s$-wave fit is about 30\% smaller.  If $\gamma_n$ were fixed at
$\gamma_{14}$ the $s$-wave fit would be considerably worse.  Clearly, optimizing $\gamma_n$ for a
fit to the low temperature data will not significantly improve the $s$-wave fit.  The values of
$\alpha$ found show that the $d$-wave coupling is rather strong.

\begin{figure}
\center
\includegraphics*[width=7cm]{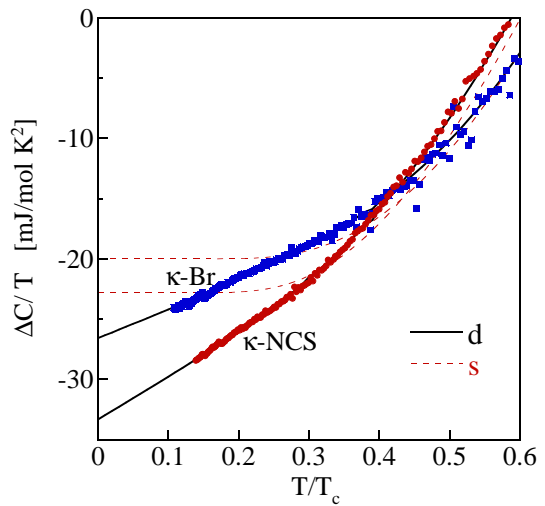}
\caption{(Color online) $\Delta C/T$ versus $(T/T_c)$ for $\kappa$-Br and $\kappa$-NCS.  The solid
lines are fits to the $d$-wave model and the dashed and dotted lines are the $s$-wave fits for each
compound respectively.} \label{figt2}
\end{figure}

The difference between the $s$-wave and $d$-wave fits is perhaps shown more clearly in Fig.\
\ref{figt2}, where we have plotted $\Delta C/T$ versus $T/T_c$. In the low temperature limit, the
clean $d$-wave model predicts $C_{el}\sim T^2$  so we expect $\Delta C/T \simeq aT-\gamma_n$. The
$s$-wave model predicts $\Delta C/T \simeq a^\prime
T^{-\frac{5}{2}}\exp(-\frac{\Delta_0}{k_BT})-\gamma_n$. The lines on the figure are the same fits
as in Figs.~\ref{figCBr} and \ref{figCNCS}. Fig.\ \ref{figt2} shows that below $T/T_c\simeq 0.3$
$\Delta C/T$ varies linearly with $T$, and the full $d$-wave model fits the data over the full
temperature range. Again, clearly the $s$-wave model does not fit the at low temperature.  A linear
fit to the $\Delta C/T$ for $T/T_c<0.3$ gives $a=2.33\pm0.03$ mJ/mol K$^3$ and $a=4.21\pm0.04$
mJ/mol K$^3$ for $\kappa$-Br and $\kappa$-NCS respectively.

The data in this paper is representative of results taken on a large number of different crystals.
In total 6 samples of $\kappa$-Br (with $T_c$ values in the range 11.5K$<T_c<$12.4K) and 3 samples
of $\kappa$-NCS (with $T_c$ values in the range 9.3K$<T_c<$9.6K) were measured and all were found
to have the same behavior as that reported here. For $\kappa$-Br it is known that fast cooling
through the temperature region 60-85K depresses $T_c$ \cite{TokumotoKTKA99}. The sample reported
here was cooled very slowly (at 0.72~K/hr) through this region. Data was also taken for higher
cooling rates, which we find significantly decreases $\gamma_n$ and $T_c$ but leaves the $T$
dependence of $C$ unchanged, except close to $T_c$. These results will be reported in detail
separately.

Within the $d$-wave model, by linearizing the angle dependence of the gap near the nodes it can be
shown that at low temperature, for a two dimensional cylindrical Fermi surface,
$\frac{C_{el}}{T^2}=\frac{54\zeta(3)k_B\gamma_n}{\pi^2\mu\Delta_0}$, where $\mu \Delta_0 =
\left.\frac{d\Delta(\phi)}{d\phi}\right|_{\rm node}$. Hence, the coefficient of the $T^2$ term in
the specific heat determines only the slope of the energy gap near the nodes, $\mu\Delta_0$. In
general, the energy gap may not simply vary like $\cos(2\phi)$ and so $\mu$ may differ from 2. We
have considered this possibility by fitting the data to a linearized gap model, where
$\Delta(\phi)=\mu\Delta_0\phi$ for $|\phi-\frac{\pi}{4}|<\frac{1}{\mu}$ and $\Delta(\phi)=\Delta_0$
otherwise.  The behavior of the data at $T/T_c \gtrsim 0.3$ constrains the values of
$\mu=2.0\pm0.4$, although clearly this depends, to some extent, on our assumed weak-coupling form
of $\Delta(T)$.

We have also considered the possibility of a mixed order parameter where the gap does not go to
zero at the `nodes', for example, $d_{x^2+y^2}+i\epsilon s$.  Our data constrains $\epsilon\lesssim
0.1 $, so the minimum gap at the `nodes' is $\lesssim$10\% of the maximum.

The low temperature ($0.3\lesssim T \lesssim 2$~K) $C_{el}$ data of Nakazawa \textit{et
al.}\cite{NakazawaK97} also follow a $T^2$ power law with the coefficient of the $T^2$ term very
similar to that reported here. However, for $T\gtrsim 2$~K their data deviate markedly from ours
presumably because of the above mentioned uncertainties in their phonon subtraction. These
uncertainties are much less important at low temperature.

Arai \textit{et al.} \cite{AraiINTYNA01} report that their tunnelling spectra  in $\kappa$-NCS can
be fitted with a $d$-wave gap with $\Delta_0$ in the range $3.0-5.7 $meV. The lower end of this
range compares reasonably well to that found here ($\Delta_0=2.6$~meV, see Table \ref{fitparams}).

Le \textit{et al.} \cite{LeLSWUBRSSYWKCW92} have reported measurements of the temperature
dependence of the superfluid density by the muon spin relaxation ($\mu$SR) technique. We have
fitted their data to the same $d$-wave model used above (again assuming a cylindrical Fermi
surface), and find $\alpha=1.7\pm0.2$ and $\alpha=1.4\pm 0.2$ for $\kappa$-Br and $\kappa$-NCS
respectively. This compares favorable to the values found here.  A quantitative analysis of the
penetration depth data of Ref.\ \cite{CarringtonBPGKSWGW99} is complicated by uncertainties in the
assumed value of the penetration depth at zero temperature, however the measured temperature
dependence of $\lambda$ is in very good agreement with predictions from the current model, once
impurity effects are included.

In conclusion, we have measured the specific heat of an extensive set of samples of $\kappa$-Br and
$\kappa$-NCS and find that in all cases the data are well fitted by a strong coupling $d$-wave
model.  Our data firmly rule out an isotropic $s$-wave gap in these samples.

Work at Argonne National Laboratory is sponsored by the U. S. Department of Energy, Office of Basic
Energy Sciences, Division of Materials Sciences, under Contract DE-AC02-06CH11357.

\end{document}